\documentstyle[prl,aps,multicol,epsf]{revtex}
\large
\begin{document}
\draft
\title{Positive and negative Hanbury-Brown and Twiss correlations 
in normal metal--superconducting devices} 

\author{Julien Torr\`es and Thierry Martin}
\address{Centre de Physique Th\'eorique,
Universit\'e de la M\'editerran\'ee,
Case 907, F-13288 Marseille Cedex 9, France}
\maketitle
\begin{abstract}
In the light of the recent analogs of the 
Hanbury--Brown and Twiss experiments \cite{Henny Oliver}
in mesoscopic beam splitters, negative current noise 
correlations are recalled to be the consequence of an 
exclusion principle. Here, positive (bosonic) correlations 
are shown to exist in a fermionic system,  
composed of a superconductor connected to two normal 
reservoirs.
In the Andreev regime, the correlations can either be 
positive or negative
depending on the reflection coefficient of the beam 
splitter. For biases beyond the gap, 
the transmission of quasiparticles favors fermionic 
correlations. The presence of disorder  
enhances positive noise correlations.
Potential experimental applications are discussed.     
\end{abstract}
\begin{multicols}{2}
\narrowtext

\pacs{PACS 74.40+k,74.50+r,72.70+m}

In condensed matter systems, correlations effects
between carriers exist either because 
particles interact with each other, or alternatively 
because the observable one considers 
involves a measurement on more than one particle. 
The characterization of current fluctuations in time
constitutes a central issue in quantum 
transport. Noise measurements have been used to 
detect the fractional charge of the excitations
in the quantum Hall effect \cite{Saminadayar,Reznikov}. 
More recently, a fermion analog of the 
Hanbury--Brown and Twiss experiment \cite{HBT}
was achieved \cite{Henny Oliver} with mesoscopic 
devices, obtaining a clear signature of the negative 
correlations expected from the Pauli principle.
Here we recall rapidly the ingredients which are 
necessary for negative correlations and we propose a 
Hanbury--Brown and Twiss experiment for a fermionic 
system were both negative and positive (bosonic) 
noise correlations can be detected. 

\begin{figure}
\epsfxsize 8.5 cm
\centerline{\epsffile{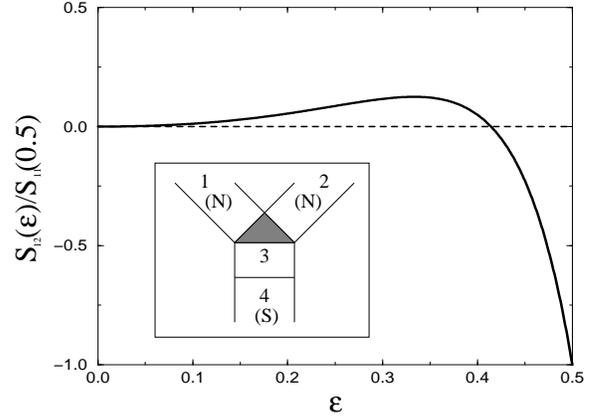}}
\medskip
\caption{\label{fig1}
Noise correlation between the two normal reservoirs of the device
(inset), as a function of the transmission probability of the 
beam splitter, showing both positive and negative correlations.
Inset: the device consists of a superconductor ($4$)--normal ($3$)
interface which is connected by a beam splitter 
(shaded triangle) to reservoirs ($1$) and ($2$)}
\end{figure}

The system which is proposed (see inset of Fig. \ref{fig1}) 
consists of a junction, or beam splitter, connected 
by electron channels to reservoirs, which 
is similar to that of Ref. \cite{Martin Landauer}, 
except that the injecting reservoir is a superconductor. 
Because of the proximity effect at the 
interface between the superconductor and the 
normal region, electrons and holes behave like 
Cooper pairs provided there is enough mixing between
them. While Cooper pairs are not bosons strictly speaking,
an arbitrary number of theses can exist in the same 
momentum state, which opens the possibility
for bosonic correlations. 
Bosonic behavior in electron systems
has been previously discussed, for instance
for excitons in coupled quantum wells \cite{Negoita},
where the possibility of observing
Bose condensation is in debate.
On the other hand, it may be possible to detect 
negative correlations in adequately prepared photonic 
systems \cite{Liu}.  

Negative noise correlations in branched electron circuits
are the consequence of an exclusion principle, which exists 
for fermions (Pauli principle) or even for particles which obey 
Haldane's exclusion statistics \cite{Haldane}. 
In these two situations, \cite{Lesovik Buttiker,Martin Landauer,Isakov},
ignoring thermal effects, low frequency current noise in a two--probe device
is suppressed by a factor $(1-T)$, with $T$ the 
transmission probability. 
Consider first a circuit with three leads
with corresponding currents and noises $I_i$ and $S_i$ 
labeled  $i=1,2,3$, as depicted in inset of Fig. \ref{fig1}
(but ignoring region $4$, which we take later to be 
a superconductor). 
Assume that particles (fermions or exclusion particles) 
with charge $q$ are injected from $3$ 
with a chemical potential $\mu_3$ while $1$ and $2$ are kept 
at the same chemical potential $\mu_1$.
Noise correlations between $1$ and $2$ can be computed by 
invoking that the fluctuations in $3$ \cite{Martin Landauer} equal 
that of a composite lead $(1+2)$: at $\omega=0$, 
$S_3=S_{(1+2)}$. The definition of the
noise correlations between $1$ and $2$ is:
\begin{equation}
S_{12}\equiv
\lim_{T \rightarrow +\infty} \frac{1}{T}
\int_{0}^{T}\int_{-\infty}^{+\infty} dt dt' 
\langle \delta I_{1}(t) \delta I_{2}(t+t') \rangle
~,\end{equation}
with $\delta I_i$ the fluctuation around the average current 
in $i$. The correlations $S_{12}$ are obtained by
subtracting the individual noise of $2$ and $3$ from $1$:
$ S_{12}=[S_3-(S_1+S_2)]/2 $. For electrons and exclusion particles, 
a multi--terminal noise formula \cite{Martin Landauer}
gives:
\begin{equation}
{S_3\over (\mu_3-\mu_1)}={2q^2\over hg}\sum_i 
Tr[\tilde{s}_{(1+2),3}\tilde{s}_{(1+2),3}^\dagger
({\bf 1}-\tilde{s}_{(1+2),3}\tilde{s}_{(1+2),3}^\dagger)]~,\end{equation}
were $\tilde{s}_{(1+2),3}$
is the ($2\times 1$) transmission matrix between $3$ and $(1+2)$,
which is a submatrix of the scattering matrix with elements
$s_{ij}$ describing the junction, ${\bf 1}$ is the 
two--dimensional identity matrix, and $g$ is the 
exclusion parameter ($g=1$ for fermions).
Using current conservation, one obtains
negative correlations for both fermion
and exclusion particles:
\begin{equation}
S_{12}=-{2q^2\over h}{(\mu_3-\mu_1)\over g}s_{13}s_{23}^\dagger 
s_{13}^\dagger s_{23}.
\end{equation}
Minimal negative correlations $S_{12}/S_1=-1$ are obtained 
for a reflectionless, symmetric junction.

Positive correlations in systems 
where the injecting lead is a superconductor are now addressed.
The scattering approach to quantum transport
in the presence of normal--superconductor (NS) boundaries
is available \cite{Beenakker Houches,Khmelnitskii,Datta}
so the basic steps are reviewed briefly.
The fermion operators which enter the current operator 
are given in terms of the quasiparticle states 
using the Bogolubov transformation \cite{BdG}
$\psi_{\sigma}(x) =\sum_n \left( u_n(x) c_{n \, \sigma} 
- \sigma v_n^*(x) c^{\dagger}_{n \, -\sigma} \right)$,
were $c^{\dagger}_{n \, \sigma}$ ($c_{n \, \sigma}$)
are quasiparticle creation (annihilation) operators,  
$n=(i,\alpha,E)$ contains information on the reservoir ($i$)
from which the particle ($\alpha=e,h$) is incident with energy $E$ 
and $\sigma$ labels the spin.
The contraction of these two operators gives
the distribution function of the particles injected from 
each reservoir, which for a potential bias
$V$ are: $f_{ie}\equiv f(E-eV)$ for electrons incoming 
from $i$, similarly $f_{ih}\equiv f(E+eV)$ for holes,
and $f_{i,\alpha}=f(E)$ for both types of quasiparticles  
injected from the superconductor ($f$ is the Fermi--Dirac distribution).
Here, $eV>0$ means that electrons are injected from
regions $1$ and $2$. Invoking
electron--hole symmetry, the (anti)correlations 
of holes are effectively studied. 
$u_n(x)$ and $v_n(x)$ are the solutions of the 
Bogolubov--de Gennes equations which contain the relevant 
information on the reflection/transmission of electrons 
and holes (and their quasiparticle analogs) at the 
NS interface. The current operator allows to derive
a general expression for the zero frequency 
noise correlations between normal terminals 
$i$ and $j$ \cite{Datta,Martin NS} 
which constitutes our starting point:
\begin{eqnarray}
\nonumber
S_{ij}(0) &=&
\frac{e^2 \hbar^2}{2m^2} \frac{1}{ 2 \pi \hbar }
\int_0^{+\infty} \!\!\! dE \sum_{\alpha,\beta}  
f_{i\alpha}(1-f_{j\beta})
\Bigl[
 A_{i \alpha j \beta}   A^*_{i \alpha j\beta}
\\
&&\label{correlations ij}
+ B^*_{i\alpha j\beta}B_{i\alpha j\beta}
+ A_{i\alpha j\beta} B_{i\alpha j\beta}
+ B^*_{i\alpha j\beta} A^*_{i\alpha j\beta}
\Bigr]
~,
\label{general correlations}
\end{eqnarray}
where current matrix elements are defined by 
$A_{i \alpha j \beta}\equiv u_{j\beta }
\partial_x u^*_{i\alpha} -  u^*_{i\alpha} \partial_x u_{j\beta}$
and $B_{i\alpha j\beta}\equiv v^*_{j\beta}\partial_x v_{i\alpha} 
- v_{i\alpha} \partial_x v^*_{j\beta}$. 
The electron and hole wave functions describing scattering states
$\alpha$ (particle) and $i$ (lead) are expressed in terms of the 
elements $s_{ij\alpha\beta}$ of the 
S--matrix which describes the whole NS ensemble:
\begin{mathletters}
\begin{eqnarray}
u_{i\alpha}(x_j)&=&[\delta_{ij}\delta_{\alpha e}e^{ik_+x_j}+
s_{jie\alpha}e^{-ik_+x_j}]/\sqrt{v_+}\\ 
v_{i\alpha}(x_j)&=&[\delta_{ij}\delta_{\alpha h}e^{-ik_-x_j}+
s_{jih\alpha}e^{ik_-x_j}]/\sqrt{v_-}
\label{u et v} 
\end{eqnarray}
\end{mathletters}
where $x_j$ denotes the position in normal lead $j$ and $k_\pm$ ($v_\pm$)
are the usual momenta (velocities) of the two branches.
$S_{ij}(0)$ has been shown to have no definite sign in four--terminal
noise measurements \cite{Datta}.

Specializing now to the NS junction connected to a beam splitter
(inset of Fig. \ref{fig1}), $6\times 6$ matrix elements are sufficient to 
describe all scattering processes. At zero temperature, 
the noise correlations between 
the two normal reservoirs simplify to:
\begin{eqnarray}
\nonumber
&&S_{12}(0)= \frac{2e^2}{h}\int_0^{eV} dE \sum_{i=1,2}\\
\nonumber
&&\times\Bigl[
  \sum_{j=1,2}
  \left( s^*_{1iee} s_{1jeh} - s^*_{1ihe} s_{1jhh} \right)
  \left( s^*_{2jeh} s_{2iee} - s^*_{2jhh} s_{2ihe} \right)
\\
\nonumber && 
~~~+\sum_{\alpha=e,h} \left( s^*_{1iee} s_{14e\alpha} - s^*_{1ihe} s_{14h\alpha} \right)
  \left( s^*_{24e\alpha} s_{2iee} - s^*_{24h\alpha} s_{2ihe} \right)\Bigr]
~,
\label{correlations splitter}
\end{eqnarray}
where the subscript $4$ denotes the superconducting lead.
The first term represents  
normal and Andreev reflection processes \cite{Andreev}, while the 
second term invokes the transmission of quasiparticles
through the NS boundary. It was noted previously
\cite{Martin NS} that in the pure Andreev regime 
the noise correlations vanish when the junction 
contains no disorder: electron (holes) incoming 
from $1$ and $2$ are simply converted into holes
(electrons) after bouncing off the NS interface. 
The central issue, whether disorder can induce
changes in the sign of the correlations, is now 
addressed. 

Consider first the pure Andreev regime, were $eV\ll \Delta$, the 
superconducting gap, for which a simple model for a disordered
NS junction \cite{Beenakker Houches} is readily available.
The junction is composed of four distinct regions (see inset Fig. 1).
The interface between $3$ (normal) and $4$ (superconductor)
exhibits only Andreev reflection, with scattering amplitude 
for electrons into holes $r_A=\gamma\exp(-i\phi)$ 
(the phase of $\gamma=\exp[-i\arccos(E/\Delta)]$ 
is the Andreev phase and $\phi$ is 
the phase of the superconductor). 
Next, $3$ is connected to two reservoirs $1$ and $2$ by a beam 
splitter which is parameterized by a single parameter 
$0<\epsilon<1/2$ identical to that of Ref. \cite{Gefen}:
the splitter is symmetric, its scattering matrix coefficients
are real, and transmission between $3$ and the reservoirs
is maximal when $\epsilon=1/2$, and vanishes at $\epsilon=0$. 
Electrons and holes undergo multiple reflections in 
central $3$ and the scattering matrix coefficient 
$s_{ij\alpha\beta}$ of the whole device are computed
using the analogy 
with a Fabry--Perot interferometer. 
\begin{eqnarray}
\nonumber
s_{ii\alpha\alpha}&=&(x-1)(1+\gamma^2 x^2)/[2(1-\gamma^2 x^2)]\\ 
\nonumber
s_{ijeh}&=&\gamma e^{-i\phi}(1-x)(1+x)/[2(1-\gamma^2 x^2)]\\
s_{ijee}&=&(x+1)(1-\gamma^2 x^2)/[2(1-\gamma^2 x^2)]~~~~i\neq j~,
\label{s Andreev}
\end{eqnarray}
were $x=\sqrt{1-2\epsilon}$ and the remaining coefficients of the 
S--matrix are found using time 
reversal symmetry.

Next one proceeds with the standard approximation $\gamma\simeq-i$ 
which applies for low biases in order to perform the energy 
integrals in Eq. 
(\ref{correlations splitter}):
\begin{equation}
S_{12}(\epsilon) = \frac{2e^2}{h} eV \frac{\varepsilon^2}{2(1-\varepsilon)^4}
\left( -\varepsilon^2 -2\varepsilon+1 \right)~. 
\end{equation}
The noise correlations vanish at $\epsilon=0$, when 
conductors $1$ and $2$ constitute a two--terminal device decoupled from the 
superconductor, and in addition, $S_{12}$ vanishes when 
$\epsilon=\sqrt{2}-1$. A plot of $S_{12}$ as a function of 
the beam splitter transmission (Fig. \ref{fig1}) indicates 
that indeed, the correlations are positive (bosonic) for 
$0<\epsilon<\sqrt{2}-1$ and negative (fermionic) for 
$\sqrt{2}-1<\epsilon<1/2$. At maximal transmission into the 
normal reservoirs ($\epsilon=1/2$), the correlations normalized
to the noise in $1$ (or $2$) give the negative minimal 
value: electrons and holes do not interfere and propagate
independently into the normal reservoirs. 
It is then expected to obtain the signature of a purely 
fermionic system. 
When the transmission $\epsilon$ is decreased, Cooper 
pairs can leak in region $3$ \cite{Abrikosov} because of 
multiple Andreev processes.
Further reducing the 
beam splitter transmission allows to balance the contribution 
of Cooper pairs with that of normal particles. 
Eq. (\ref{correlations splitter}) predicts maximal (positive)
correlations at $\epsilon=1/3$: a compromise
between a high density of Cooper pairs and weak transmission.  

The model described above may not be convincing
enough, as an ideal Andreev interface was assumed. 
Moreover, it
does not allow to generalize the results to the case where 
quasiparticles in the superconductor contribute to the 
current. Quasiparticles have fermionic statistics, so  
their presence is expected to cancel the positive 
contribution of Cooper pairs leaking on the normal side.
In particular, in the limit were $eV\gg \Delta$,
one should recover fermionic correlations. 

These issues bear similarities with a recent discussion of 
singularities in the finite frequency noise of 
NS junctions \cite{Lesovik Martin Torres}: below the gap,
a singularity exists at the Josephson frequency, while 
above gap there appear additional features at 
$eV\pm\Delta$ associated
with electron and hole--like quasiparticles, 
which give single--particle behavior 
in the limit $eV\gg \Delta$. However, finite frequency noise
probes the charge ($e$ or $2e$) of the carriers, 
while here one is probing the statistics of the 
effective carriers in the junction. 

The energy dependence 
of the scattering coefficients is therefore needed 
to describe the correlations away from the pure 
Andreev regime. The numerical calculations which 
follow are performed using the BTK model 
\cite{Blonder}, where the disordered interface
between regions $3$ and $4$ (inset of Fig. \ref{fig1})
can be characterized by a small number of parameters: 
the pair amplitude is assumed to be a step function 
$\Delta(x)=\Delta \Theta(x)$ and a delta function 
potential barrier $V(x)=\hbar v_FZ\delta(x)$ is imposed,
where $v_F$ the Fermi velocity and $Z\gg 1$ ($Z\ll 1$) for 
strong (weak) disorder. The beam splitter is taken to be 
similar to the previous calculation \cite{Gefen},
assuming that the 
reflection/transmission of electrons does not depend 
significantly on the incoming energy.    
   
\begin{figure}
\epsfxsize 8.5 cm
\centerline{\epsffile{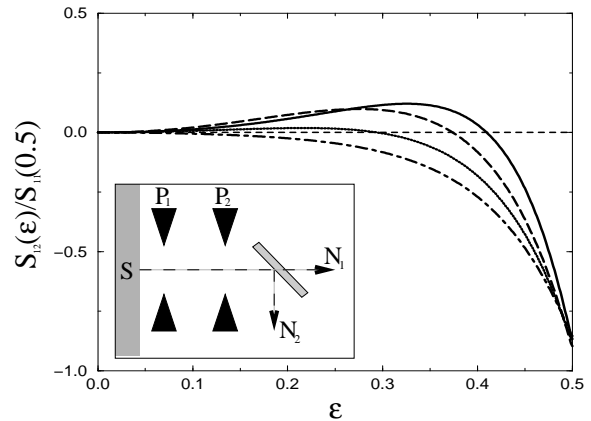}}
\medskip
\caption{\label{fig2}
Noise correlations using an NS boundary modeled by BTK
for weak disorder, $Z=0.1$: from top to bottom 
$eV/\Delta=0.5,0.95,1.2,1.8$. Inset: proposed device for the 
observation of positive/negative correlations; 
at the boundary of a superconductor ($S$), two point 
contacts ($P_1$) and ($P_2$) are connected to a 
semitransparent mirror ($M$)}
\end{figure}

Consider the case of weak disorder, $Z=0.1$ (Fig. 2). 
At weak biases, good agreement is found with the previous 
analytical results displayed in Fig. \ref{fig1}, except 
that for a fully transmitting splitter, the ratio of the 
correlations divided by the noise in region $1$ does not reach 
the extremal value $-1$:
an early signature of disorder. 
When the bias is further increased but is kept below the gap,
the phase accumulated in Andreev processes by electrons 
and holes with various energies is spread over the interval
$[0,\pi/2]$: positive correlations are weaker, but they 
survive at low beam splitter transmission. 
Further increasing the voltage beyond the gap destroys 
completely the bosonic signature of the noise. 

\begin{figure}
\epsfxsize 8.5 cm
\centerline{\epsffile{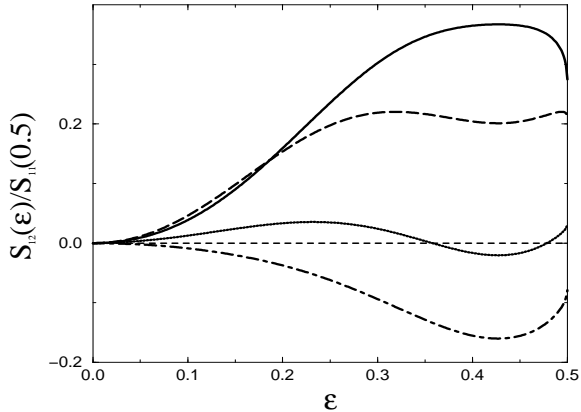}}
\medskip
\caption{\label{fig3}
Noise correlations using an NS boundary modeled by BTK
for intermediate disorder:
$Z=1$ (same biases as in Fig. 2)}
\end{figure}

A strikingly different behavior is obtained 
for intermediate disorder at $Z=1$ (Fig. \ref{fig3}). 
First, for weak biases, the noise correlations remain 
positive over the whole range of $\epsilon$, with
a maximum located at $\epsilon \simeq 0.43$, which is 
close to the case of a reflectionless splitter. 
This maximum becomes a local minimum for higher biases,
where positive correlations remain quite robust 
nevertheless. Just below the gap ($eV=0.95$), correlations
oscillate between the positive and the negative sign,
but further increasing the bias eventually favors 
a fermionic behavior. Calculations for larger values of 
$Z$ confirm the tendency of the system towards
dominant positive correlations at low biases
with $S_{12}(\epsilon)/S_1(1/2)> 1$ over a wide range of 
$\epsilon$ (not shown). 
The phenomenon of positive correlations in fermionic 
systems with a superconducting injector is thus 
{\it enhanced} by disorder at the NS boundary. 
Nevertheless, for strong disorder,
the absolute magnitude of $S_{1}$ and $S_{12}$
becomes rather small, which limits the possibility of an
experimental check in this regime.
 
An interesting feature of the present results 
is the fact that both positive and negative 
correlations are achieved in the same system.
A suggestion for this device is depicted 
in the inset of Fig. 2. Assume that a 
high mobility two dimensional electron gas has
a rather clean interface with a superconductor
\cite{van Wees}. 
A first point contact close to the 
interface selects a maximally occupied electron channel,
which is incident on a semi--transparent mirror similar
to the one used in the Hanbury--Brown and Twiss
fermion analogs \cite{Henny Oliver}. A second point 
contact located in front of the mirror, allows 
to modulate the reflection of the splitter in order 
to monitor both bosonic and fermionic noise correlations. 
In addition, by choosing a superconductor with 
a relatively small gap, 
one could observe the dependence of the
correlations on the voltage bias without encountering 
heating effects in the normal metal. 

Hanbury--Brown and Twiss type experiments may become 
a useful tool to study statistical effects in mesoscopic devices.
Here, noise correlations have been shown to have
either a positive or a negative sign in the same system.  
Close to the boundary,
a fraction of electrons and holes are correlated. 
This can be viewed as a finite density of Cooper pairs 
which behave like bosons. 
The presence of disorder allows in some cases to 
enhance the appearance of bosonic correlations. Similar 
studies could be envisioned in the Fractional 
Quantum Hall Effect (FQHE) 
where the collective 
excitations of the correlated electron fluid 
have unconventional statistics \cite{Haldane}. 
  
Discussions with the late R. Landauer, with D.C. Glattli
and with M. Devoret
are greatfully acknowledged. 

\end{multicols}

\begin{thebibliography}{99}
%
\bibitem{Henny Oliver} M. Henny {\it et al.}, Science {\bf 284}, 296 (1999);
W. Oliver {\it et al.}, {\it ibid}, 299 (1999).
%
\bibitem{HBT} R. Hanbury--Brown and Q. R. Twiss, Nature {\bf 177}, 27 (1956).
%
\bibitem{Saminadayar} 
L.\ Saminadayar, D.~C.\ Glattli, Y.\ Jin, and B.\ Etienne,
Phys.\ Rev.\ Lett.\ {\bf 79}, 2526 (1997).  
%
\bibitem{Reznikov} 
R.~de-Picciotto, M.~Reznikov, M.~Heiblum, V.\ Umansky, G.\ 
Bunin, and D.\ Mahalu, Nature {\bf 389}, 162 (1997).
\bibitem{Gefen} Y. Gefen, Y. Imry and R. Landauer, Phys. Rev. Lett. 
{\bf 52}, 139 (1984).
%
\bibitem{Martin Landauer} T. Martin and R. Landauer, Phys. Rev. B {\bf 45},
1742 (1992). 
%
\bibitem{Negoita} V. Negoita, D. W. Snoke and K. Eberl, Cond. Mat. 9901088
(and references therein).
%
\bibitem{Liu} R. Liu, private communication.
%
\bibitem{Haldane} F.D.M. Haldane, Phys. Rev. Lett. {\bf 67}, 937 (1991).
%
\bibitem{Lesovik Buttiker} G.B. Lesovik, JETP Lett. {\bf 49}, 592 (1989);
M. B\"uttiker, Phys. Rev. Lett. {\bf 65}, 2901 (1990).
%
\bibitem{Isakov} S. Isakov, T. Martin and S. Ouvry, 
Cond. Mat. 9811391. 
%
\bibitem{Beenakker Houches} C.W.J. Beenakker, in {\it Mesoscopic 
Quantum Physics}, eds. E. Akkermans {\it et al.}, p. 279 
(Les Houches LXI, North Holland 1995).
%
\bibitem{Khmelnitskii} B.~A.\ Muzykantskii and D.~E.\ Khmelnitskii, 
Phys.\ Rev.\ B {\bf 50}, 3982 (1994).
%
\bibitem{Datta}M.~P.~Anantram and S.~Datta, Phys.\ Rev.\ B {\bf 53}, 16 390 
(1996).
%
\bibitem{BdG} P.G. de Gennes, {\it Superconductivity of Metals and Alloys}, (Addison Wesley, 1966, 1989). 
%
\bibitem{Martin NS} T. Martin, Phys. Lett. A {\bf 220}, 137 (1996). 
%
\bibitem{Andreev} A.F. Andreev, J. Exp. Theor. Phys.
{\bf 46}, 1823 (1964) [Sov. Phys. JETP {\bf 19}, 1228 (1964)].
%
\bibitem{Abrikosov} See for example, A.A. Abrikosov, {\it Fundamentals of the Theory of Metals}, (North-Holland, 1988)
%
\bibitem{Lesovik Martin Torres} 
G. Lesovik, T. Martin and J. Torr\`es, Cond. Mat. 9902278; 
J. Torr\`es, G. Lesovik and T. Martin, ({\it in preparation}, 1999). 
%
\bibitem{Blonder} G.E. Blonder, M. Tinkham and T.M. Klapwijk, 
Phys. Rev. B {\bf 25}, 4515 (1982). 
%
\bibitem{van Wees} A. Dimoulas {\it et al.}, Phys. Rev. Lett. {\bf 74}, 602 (1995).
%
%
\end{thebibliography}
\end{document}